\documentclass[%
reprint,
superscriptaddress,
amsmath,amssymb,
prx,
floatfix,
]{revtex4-2}

\usepackage{color}
\usepackage{xcolor}
\usepackage[separate-uncertainty = true,multi-part-units=single]{siunitx}
\usepackage{float}
\usepackage{xspace}
\usepackage{natbib}
\usepackage{multirow}
\usepackage{hhline}
\usepackage{graphicx}% Include figure files
\usepackage{dcolumn}% Align table columns on decimal point
\usepackage{bm}% bold math
\usepackage{color}
\usepackage{hyperref}
\newcommand{\singlecolumn}{86mm}
\newcommand{\doublecolumn}{176mm}

\newcommand{\singlet}{$\left| \textrm{S} \right>$\xspace} % for singlet ket
\newcommand{\tzero}{$\left| \textrm{T}_0 \right>$\xspace} % for T0 ket
\newcommand{\tplus}{$\left| \textrm{T}_+ \right>$\xspace} % for T+ ket
\newcommand{\tminus}{$\left| \textrm{T}_- \right>$\xspace} % for T- ket
\newcommand{\tone}{$\textrm{T}_1$\xspace} % for T1

\newcommand{\Btwo}{$\textrm{B}_{\textrm{2}}$\xspace}
\newcommand{\Bthree}{$\textrm{B}_{\textrm{3}}$\xspace}

\newcommand{\updown}{$\left| \uparrow\mathrel{\mspace{-1mu}}\downarrow \right>$\xspace} % for spin states
\newcommand{\downup}{$\left| \downarrow\mathrel{\mspace{-1mu}}\uparrow \right>$\xspace} % for spin states
\newcommand{\Sotwo}{$\left| S (0,2) \right>$\xspace} % for singlet ket
\newcommand{\Soneone}{$\left| S (1,1) \right>$\xspace} % for singlet ket
\newcommand{\Toneone}{$\left| T_0 (1,1) \right>$\xspace} % for singlet ket
\newcommand{\Totwo}{$\left| T_0 (0,2) \right>$\xspace} % for singlet ket

\newcommand{\Sotwoexc}{$\left| S_1 (0,2) \right>$\xspace} % for singlet ket

\begin{document}
%>

%< Authors and affiliations
\title{Complete readout of two-electron spin states in a double quantum dot}

\author{Martin Nurizzo}
\affiliation{Univ. Grenoble Alpes, CNRS, Grenoble INP, Institut N\'eel, F-38000 Grenoble, France}%
% martin.nurizzo@neel.cnrs.fr

\author{Baptiste Jadot}
\affiliation{Univ. Grenoble Alpes, CEA, Leti, F-38000 Grenoble, France}%
% baptiste.jadot@cea.fr

\author{Pierre-Andr\'e Mortemousque}
\affiliation{Univ. Grenoble Alpes, CEA, Leti, F-38000 Grenoble, France}%
% pierre-andre.mortemousque@cea.fr

\author{Vivien Thiney}
\affiliation{Univ. Grenoble Alpes, CEA, Leti, F-38000 Grenoble, France}%
% vivien.thiney@neel.cnrs.fr

\author{Emmanuel Chanrion}
\affiliation{Univ. Grenoble Alpes, CNRS, Grenoble INP, Institut N\'eel, F-38000 Grenoble, France}%
% emmanuel.chanrion@neel.cnrs.Fr

\author{David Niegemann}
\affiliation{Univ. Grenoble Alpes, CNRS, Grenoble INP, Institut N\'eel, F-38000 Grenoble, France}%
% david.niegemann@neel.cnrs.Fr

\author{Matthieu Dartiailh}
\affiliation{Univ. Grenoble Alpes, CNRS, Grenoble INP, Institut N\'eel, F-38000 Grenoble, France}%
% matthieu.dartiailh@neel.cnrs.Fr

\author{Arne Ludwig}
\affiliation{Lehrstuhl f{\"u}r Angewandte Festk{\"o}rperphysik, Ruhr-Universit{\"a}t Bochum, Universit{\"a}tsstra{\ss}e 150, D-44780 Bochum, Germany}%
% Arne.Ludwig@rub.de

\author{Andreas D. Wieck}
\affiliation{Lehrstuhl f{\"u}r Angewandte Festk{\"o}rperphysik, Ruhr-Universit{\"a}t Bochum, Universit{\"a}tsstra{\ss}e 150, D-44780 Bochum, Germany}%
% andreas.wieck@rub.de

\author{Christopher B{\"a}uerle}
\affiliation{Univ. Grenoble Alpes, CNRS, Grenoble INP, Institut N\'eel, F-38000 Grenoble, France}%
% matias.urdampilleta@neel.cnrs.fr

\author{Matias Urdampilleta}
\affiliation{Univ. Grenoble Alpes, CNRS, Grenoble INP, Institut N\'eel, F-38000 Grenoble, France}%
% matias.urdampilleta@neel.cnrs.fr

\author{Tristan Meunier}
\affiliation{Univ. Grenoble Alpes, CNRS, Grenoble INP, Institut N\'eel, F-38000 Grenoble, France}%
% tristan.meunier@neel.cnrs.fr

%\affil[*]{tristan.meunier@neel.cnrs.fr}
%\affil[+]{pierre-andre.mortemousque@neel.cnrs.fr}
% \affil[+]{these authors contributed equally to this work}
%\keywords{Keyword1, Keyword2, Keyword3}
%>

%< Abstract
\begin{abstract}
	We propose and demonstrate complete spin state readout of a two-electron system in a double quantum dot probed by an electrometer.
	The protocol is based on repetitive single shot measurements using Pauli spin blockade and our ability to tune on fast timescales the detuning and the interdot tunnel coupling between the \si{\giga\hertz} and sub-\si{\hertz} regime.
	A sequence of three distinct manipulations and measurements allows establishing if the spins are in $S$, $T_0$, $T_+$ or $T_-$ state.
	This work points at a procedure to reduce the overhead for spin readout, an important challenge for scaling up spin qubit platforms.
\end{abstract}

\maketitle
%>

%< Introduction
\section{Introduction}

Pauli principle plays a central role in the functioning of individual electron spin qubits in semiconductor quantum dot (QD) arrays \cite{boterSpiderwebArraySparse2021,philipsUniversalControlSixqubit2022b}.
For rapid and high-fidelity spin readout, it is required to convert the spin information into distinct charge configurations.
To implement it, the basic building block is a tunnel coupled double quantum dot (DQD) system filled with an electron acting as a qubit and one as an ancilla and probed via a close-by electrometer or RF-gate reflectometry \cite{urdampilletaGatebasedHighFidelity2019,betzDispersivelyDetectedPauli2015}.
Obtaining the full spin information of the two-electron system in this simple unit cell requires nevertheless extra hardware such as QDs or electron reservoirs, that represents an important overhead for future scaling of the platform \cite{nowackSingleShotCorrelationsTwoQubit2011}.
In this article, we propose and demonstrate a strategy to perform complete spin state readout of a two-electron system with a minimal footprint.
It is based on the repetition of three single shot measurements using Pauli spin blockade and our recent development in the control of the potential detuning ($\epsilon$) at the \si{\ns} timescale and of the tunnel coupling ($t_c$) between the \si{\giga\hertz} and the sub-\si{\hertz} regime.
It allows us to completely separate in the $\epsilon$, $t_c$ parameter space the two important processes of the readout: fast spin to charge conversion and the charge configuration readout in the sub-\si{\hertz} interdot tunnel coupling regime where the system is unaffected by charge state relaxation.
Doing so, the three basic measurement procedures and their repeatability are investigated: a high fidelity $S$-$T$ readout ($F_{\textrm{PSB}} = \SI{98.43}{\percent}$) preserving the initial spin state is demonstrated.
Second, via precise navigation in the $\epsilon$, $t_c$ parameter space, we engineered and characterized an on-demand parity readout procedure relying on a selective relaxation hotspot of $T_0$ to $S$ and reached a fidelity of $F_{\textrm{parity}} = \SI{93.3}{\percent}$.
We then study the selective transformation of $T_+$ to $S$ via an adiabatic passage through the avoided crossing and demonstrate a transformation fidelity of \SI{79.9}{\percent}.
Finally, we characterize the complete readout procedure  allowing to discriminate a $S$, $T_0$, $T_+$ and $T_-$ spin state by interleaving spin readout and manipulations on the two same electrons.
This complete readout procedure is implemented for three different initializations and in more complex spin manipulation sequences such as exchange controlled oscillations to test its validity.
%>

%< Frozen Pauli Spin Blockade in the isolated regime
\section{Frozen Pauli spin blockade, a high fidelity repeatable readout}
\label{sec:FPSB}

%<< Figure : PSB
\begin{figure}[h!]
	\centering
	\includegraphics[width=\singlecolumn]{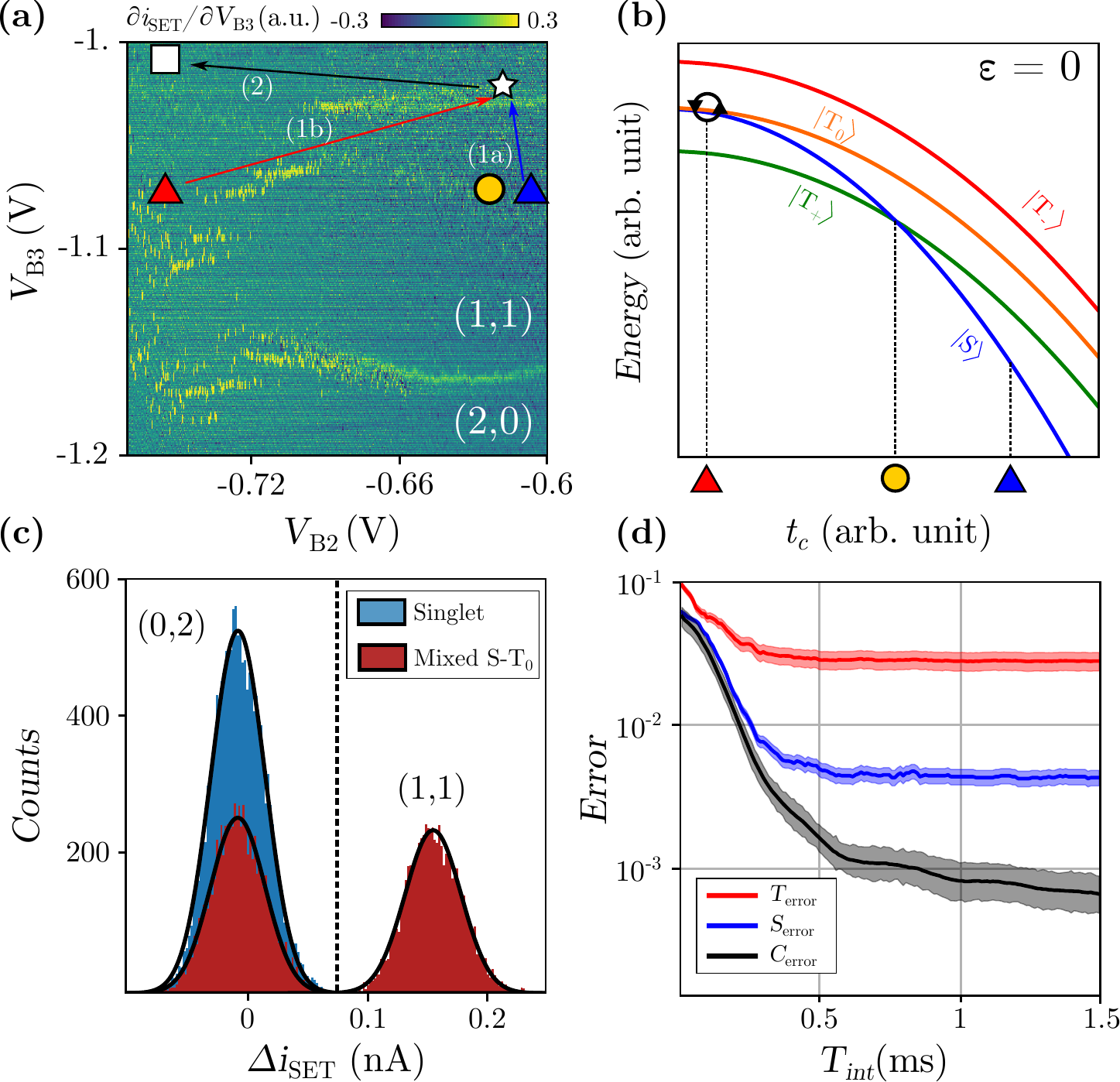}
	\caption{
		\textbf{Frozen Pauli spin blockade measurement: protocol and fidelity.}
		(a)~Stability diagram and frozen Pauli spin blockade measurement scheme.
		The voltage $V_{\textrm{B2}}$ controls mostly the tunnel coupling of the DQD while $V_{\textrm{B3}}$ controls $\epsilon$.
		The tunnel coupling can be controlled from the sub-\si{\hertz} to \si{\giga\hertz} regime in a few hundreds of \si{\milli\volt} applied to \Btwo gate.
		(b)~Sketch of the energy diagram of the relevant spin states of two electrons in a DQD.
		At the blue triangle the exchange energy in between the two spins is high enough to preserve the singlet state.
		At the red triangle the exchange energy is low compared to the magnetic field gradient and mixing of the initial S spin state occurs with the $T_0$.
		At the yellow circle, the $S$ and $T_+$ states are degenerated and mixing occurs via the Overhauser magnetic field and spin orbit coupling.
		(c)~Histogram of \num{50000} single shot measurements with an integration time of $T_{\textrm{int}}=\SI{1.5}{\ms}$.
		Following Connors et al. \cite{connorsRapidHighFidelitySpinState2020a} we define a threshold (dashed line) to discriminate the (1,1) and (0,2) charge state minimizing the readout error.
		(d)~Fidelity of the frozen PSB readout procedure as a function of the electrometer signal integration time.
		For $T_{int}<\SI{0.5}{\milli\second}$, the spin readout error is limited by the charge readout error $C_{\textrm{error}}$.
		For longer integration times the spin readout error saturates and reaches $S_{\textrm{error}} = \SI{0.43(5)}{\percent}$ and $T_{\textrm{error}} = \SI{2.7(4)}{\percent}$.
		The uncertainty is represented by the shade around the solid line.
	}
	\label{fig:FigPSB}
\end{figure}
%>> Figure : PSB

In the paradigm of spin qubits stored in gate defined QDs, the readout fidelity via classic PSB scheme is limited by the relaxation at the measurement position \cite{barthelRelaxationReadoutVisibility2012,barthelRapidSingleShotMeasurement2009, connorsRapidHighFidelitySpinState2020a,urdampilletaGatebasedHighFidelity2019}.
To overcome this issue several methods have been developed to reduce the time needed to discriminate with a high fidelity the possible charge configurations mapping $T$ and $S$ spin states.
Latching- and cascade-based mechanisms recently allowed an enhancement of the output signal of the spin to charge conversion mechanisms, improving the spin readout efficiency for an identical noise and integration time \cite{harvey-collardHighFidelitySingleShotReadout2018,vandiepenElectronCascadeDistant2021,nakajimaRobustSingleShotSpin2017}.
However, at the end of the readout procedure the initial spin state is destroyed due to the presence of an extra electron or its fast displacement in the array, meaning that the readout is not repeatable.
To tackle this issue, we implemented an original method entitled frozen Pauli spin blockade (FPSB) which consists in performing the spin to charge conversion and the charge configuration readout at two different positions in the $\epsilon$ and $t_c$ parameter space.
The spin to charge mapping is obtained in the \si{\giga\hertz} tunnel coupling ($t_c$) regime at the classic PSB position where charge transfer from (1,1) to (0,2) is allowed only for the $S$ state.
Once obtained, the charge dynamic is blocked by setting $t_c$ in the sub-\si{\hertz} regime \cite{nurizzoControlledQuantumDot2022b} to prevent any subsequent alteration of the obtained charge distribution, in particular due to spin relaxation.

To implement this readout we first characterized the DQD presented in Fig.~\ref{fig:SUP-device} by performing a stability diagram with two electrons loaded in the system \cite{mortemousqueCoherentControlIndividual2021,nurizzoControlledQuantumDot2022b}.
The stability diagram is performed as a function of the voltage gates \Btwo and \Bthree controlling respectively $t_c$ and $\epsilon$.
The result is plotted in Fig.~\ref{fig:FigPSB}(a) where we observe two discontinuous charge degeneracy lines separating the three possible charge states of two electrons in a DQD (2,0), (1,1), (0,2).
Navigating along one of the charge degeneracy line we identify two possible regimes of the inter-dot tunnel coupling.
The first one is the high tunnel coupling regime ($t_c \simeq \si{\giga\hertz}$) obtained for $V_{\textrm{B2}} > \SI{-0.68}{\volt}$.
In this regime electrons can be transferred from one dot to the other by varying $V_{\textrm{B3}}$.
For $V_{\textrm{B2}} < \SI[]{-0.68}{\volt}$, the charge degeneracy line is shifted to lower values of $V_{\textrm{B3}}$ indicating that the inter-dot tunnel rate is lower than the measurement sweep rate, in this case \SI{250}{\milli\volt\per\second} \cite{bertrandQuantumManipulationTwoElectron2015a,eeninkTunableCouplingIsolation2019a,nurizzoControlledQuantumDot2022b}.
This diagram demonstrates our ability to initialize the DQD with any charge configuration and tune $t_c$ over several orders of magnitude \cite{nurizzoControlledQuantumDot2022b}.
We will now see how navigating in the voltage gate space allows us to implement the FPSB spin measurement protocol.

The FPSB readout procedure relies on two operating positions indicated on top of the stability diagram.
The first one is the PSB position at the white star close to the (1,1)/(0,2) charge transition where the spin to charge conversion process is performed.
And the charge readout position (indicated by the white square) located in the sub-{\si{\hertz}} inter-dot tunnel coupling regime where the signal of the electrometer is acquired.
To perform the FPSB readout we will therefore use a two-pulse sequence: first the system is pulsed to the PSB position during a time long enough to allow charge transfer in the case of a $S$ state but short compared to the $\textrm{T}_1$ relaxation ($\sim\SI{10}{\micro\second}$), before being pulsed to the fixed charge readout position where $\textrm{T}_1 \gg \SI{1}{\second}$ and where sensing has been optimized.

For the demonstration and calibration of the readout two initializations of $S$ and $T$ are necessary \cite{jadotDistantSpinEntanglement2021}.
To do so, we used the high level of control over the tunnel coupling to initialize either a $S$ state or a so called $S$-$T_0$ mixed state composed of \SI{50}{\percent} $S$ and \SI{50}{\percent} $T_0$.
We engineered and implemented the pulse sequence sketched on top of the stability diagram in Fig.~\ref{fig:FigPSB}(a).
First, two electrons are loaded in the (0,2) charge configuration and we wait \SI{5}{\ms} for the electrons to relax to $S$.
From there, a \SI{500}{\ns} pulse to the blue or red triangle position initializes a $S$ or $S$-$T_0$ mixed state respectively.
As we can see in the energy diagram plotted in Fig.~\ref{fig:FigPSB}(b), at the blue triangle position $t_c$ is large enough so that the ground state of the system remains \Soneone.
On the contrary, at the red triangle the low inter-dot tunnel coupling and the random transverse magnetic field gradient allows spin mixing between the $S$ and $T_0$ state resulting in a statistical mixture of \SI{50}{\percent} $S$ and \SI{50}{\percent} $T_0$ \cite{merkulovElectronSpinRelaxation2002,pettaCoherentManipulationCoupled2005a,jadotDistantSpinEntanglement2021}.
Once the desired spin state is initialized the system is pulsed during \SI{20}{\nano\second} at the PSB position (white star in Fig.~\ref{fig:FigPSB}) to perform the spin to charge conversion.
At this position, a $S$ state is mapped to a (0,2) and $T_0$ , $T_+$ , $T_-$ are mapped to a (1,1) charge state.
The system is then pulsed to the white square, via a \si{ns}-pulse, where charge tunneling is no longer possible at the timescale of the experiment.
At this position the signal of the electrometer $\Delta i_{\textrm{SET}}$ is integrated during a time $T_{\textrm{int}}$.
We performed the measurement sequence for both initializations \num{50000} times and plotted the histogram of the integrated electrometer signal in Fig.~\ref{fig:FigPSB}(c).
We fit the histogram with a sum of two Gaussian distributions for the possible two charge states.
As expected the singlet initialization yields almost for each measurement shots a (0,2) output, while we observe an equipartition of (1,1) and (0,2) for the $S$-$T_0$ mixed state one.
Through this experiment we demonstrate the feasibility of separating the spin to charge conversion process from the charge readout in the voltage gate space while still being able to perform a high fidelity spin readout.

We can now benchmark the readout procedure using the two deterministic initializations presented previously.
In this readout two types of errors are possible, $T_{\textrm{error}}$ the probability to measure (0,2) when the spins are in one of the triplet state and $S_{\textrm{error}}$ the probability to measure (1,1) when the system is in singlet.
To assess both of these errors we used the two initializations presented and the following equations:

\begin{eqnarray}
	P_{(0,2)}^{\textrm{S}} &=& 1-S_{\textrm{error}}\;,
	\\
	P_{(0,2)}^{\textrm{mixed}} &=& \frac{1}{2}-\frac{S_{\textrm{error}}+T_{\textrm{error}}}{2} \;.
\end{eqnarray}

Where $P_{(0,2)}^{\textrm{S}}$ and $P_{(0,2)}^{\textrm{mixed}}$ are the probabilities to readout a (0,2) charge state when the system is initialized in $S$ and $S$-$T_0$ mixed state respectively.
Finally, the charge readout error $C_{\textrm{error}}$ is defined by fitting the overlap of the two histograms in Fig.~\ref{fig:FigPSB}(c) \cite{connorsRapidHighFidelitySpinState2020a}.
The three types of errors are acquired for a variable $\Delta i_{\textrm{SET}}$ integration time $T_{\textrm{int}}$ and averaged over \num{50000} measurement shots in Fig.~\ref{fig:FigPSB}(e).
For low integration time $T_{\textrm{int}}<\SI{0.5}{\milli\second}$, the spin readout error is limited by the ability to discriminate (1,1) from (0,2) charge state.
For longer integration time, the charge readout error keeps decreasing but the $S_{\textrm{error}}$ and $T_{\textrm{error}}$ saturates at respectively \SI{0.43(5)}{\percent} and \SI{2.7(4)}{\percent}.
This method to assess the fidelity gives us a higher bound about the readout errors since the initialization errors cannot be properly distinguished from it.
We explain the $S_{\textrm{error}}$ value by considering the thermal population of the triplet state at the initialization position yielding an electronic temperature of $T_{\textrm{el}} = \SI{200}{\milli\kelvin}$ close to previously measured temperatures in a similar configuration.
By spending only \SI{20}{\nano\second} at the PSB position most of the errors due to the \tone relaxation are suppressed, and the spin readout error is not increasing for higher $T_{\textrm{int}}$.
We therefore attribute the residual error value mostly to the readout calibration (see Fig.~\ref{fig:SUP-PSB-calibration}).
In the end, the separation of the spin to charge conversion and the charge readout allows for a higher charge readout fidelity without any compromise on the spin to charge conversion fidelity \cite{urdampilletaGatebasedHighFidelity2019,connorsRapidHighFidelitySpinState2020a}.

%> Electron isolation from the reservoir

%< Parity readout
\section{$T_0$ selective relaxation}

%>> Figures/parity
\begin{figure}[h!]
	\centering
	\includegraphics[width=\singlecolumn]{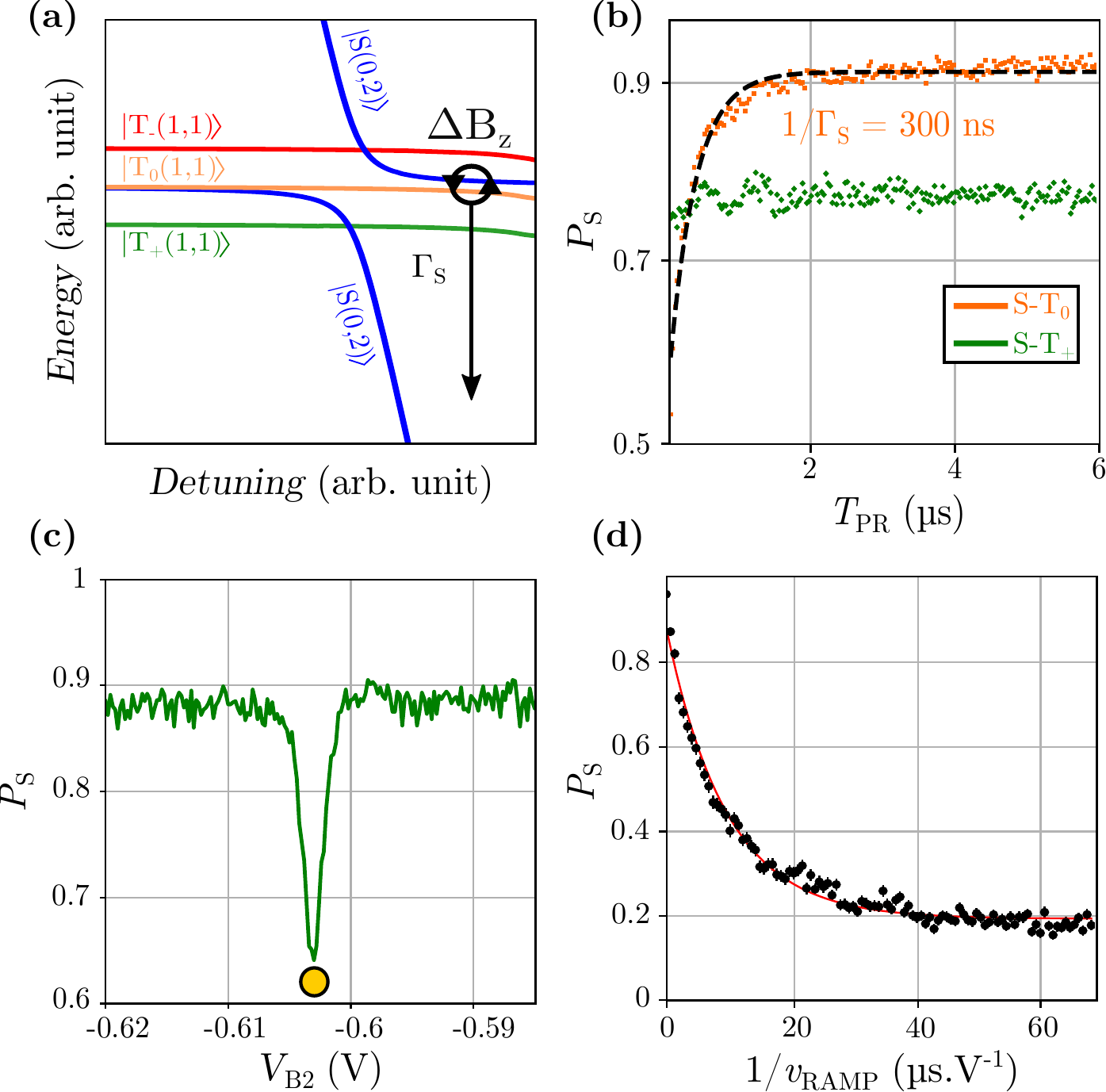}
	\caption{
		\textbf{Tools for the complete readout procedure : parity readout via $T_0$ relaxation hotspot, $S$-$T_+$ adiabatic transformation.}
		(a)~Energy diagram of the different spin states of two electrons in a DQD as function of $\epsilon$.
		The fast relaxation channel of the $T_0$ is indicated via the black arrow.
		(b)~Selective relaxation at the parity readout hotspot.
		The system is initialized in either a $S$-$T_0$ or $S$-$T_+$ mixed state and then pulsed during a variable time $T_{\textrm{HS}}$ at the parity readout position.
		(c)~Localization of the $S$-$T_+$ avoided crossing in the voltage gate space.
		(d)~Singlet return probability after a ramp through the $S$-$T_+$ avoided crossing.
	}
	\label{fig:tools}
\end{figure}
%<< Figures/parity

Classically, a PSB measurement allows to discriminate a $S$ state from $T_0$, $T_+$ and $T_-$.
However, it has been demonstrated that the relaxation of the $T_0$ state to $S$ can be strongly enhanced via either spin orbit coupling or spin state mixing by setting the system at a particular position in the parameter space close to the PSB \cite{barthelRelaxationReadoutVisibility2012,seedhousePauliBlockadeSilicon2021a,yangOperationSiliconQuantum2020,niegemannParitySinglettripletHigh2022a}.
A spin readout performed in such conditions will give the same signal for a $S$ and $T_0$ and is referred as a parity readout.

The PSB position is located close to the (1,1)/(0,2) charge transition, as we can see in the energy diagram in Fig.~\ref{fig:tools}(a), in this region the \Toneone state is isolated from the other states energy-wise.
At this position the relaxation of such state to \Sotwo is close to a few tens of \si{\micro\second} \cite[]{barthelRapidSingleShotMeasurement2009,barthelRelaxationReadoutVisibility2012}.
However, this relaxation process can be strongly enhanced by setting the system at the position in the $\epsilon$ and $t_c$ parameter space where the \Toneone and \Soneone states are nearly degenerated (see Fig.~\ref{fig:SUP-parity-hotspot}).
At this position, indicated by the circular arrows in Fig.~\ref{fig:tools}(a), spin mixing between the two states occurs in a few tens of \si[]{\ns} via the Overhauser magnetic field gradient.
The initial $T_0$ is therefore transformed in \Soneone state which rapidly relaxes to \Sotwo via phonon emission \cite{barthelRelaxationReadoutVisibility2012,meunierExperimentalSignaturePhononmediated2007}.

We demonstrate this selective relaxation enhancement of the $T_0$ state by performing the following spin manipulation procedure.
First, the spin state is initialized in a mixture of $S$-$T_0$ via the sequence presented in the previous section.
The system is then pulsed to the parity readout position described previously during a variable time $T_{\textrm{PR}}$ and finally the $S$ state probability is measured via the FPSB protocol.
In Fig.~\ref{fig:tools}(b), we plot the singlet probability $P_{\textrm{S}}$ as a function of the time spent at the parity readout position.
In the case of the unpolarized spin state initialization, we observe a rapid relaxation of the $T$ portion of the signal to the $S$ state with a characteristic rate of $1/\Gamma_{\textrm{S}} = \SI[]{300}{\ns}$.
This fast relaxation is the signature of a $T_0$ relaxation hotspot that can be used to perform a parity readout as described previously.
To ensure the $T_0$ state selectivity of the relaxation hotspot, we repeat the same procedure for a $S$-$T_+$ mixture obtained via a \SI[]{500}{\ns} pulse at the $S$-$T_+$ avoided crossing (yellow circle in Fig.~\ref{fig:FigPSB}(a,b) and \ref{fig:tools}(c)).
We observe in this case the conservation of the $T$ proportion of the signal during the whole duration spend at the parity readout position.
Indeed, the $T_+$ relaxation is not enhanced at this position and no relaxation is observed which demonstrates the selectivity of the relaxation hotspot.

\section{$S$-$T_+$ adiabatic transformation}

Having presented the singlet and parity readout mechanisms, the last tool we need is selective and rapid conversion from $T_+$ to $S$.
This is done using an adiabatic passage through the avoided crossing of the two spin states \cite{pettaCoherentBeamSplitter2010a}.
In order to perform such transformation, it is necessary to locate the avoided crossing in the voltage gate space by pulsing the tunnel coupling and observing the loss of $S$ population due to mixing when $S$ and $T_+$ are degenerated.
The spin state is initialized in a $S$ state at a large value of tunnel coupling in the (1,1) charge region, from there the tunnel coupling is pulsed during \SI[]{500}{\ns} via a voltage pulse of amplitude $V_{\textrm{B2}}$.
Finally, the spin state is read using a FPSB measurement sequence and we plot in Fig.~\ref{fig:tools}(c) the singlet probability as a function of the pulse amplitude.
A sharp drop of $P_{S}$ is observed for $V_{\textrm{B2}} = \SI[]{-0.602}{\volt}$ and corresponds to the system being pulsed to the $S$-$T_+$ avoided crossing.
At this position state mixing occurs through either the spin-orbit coupling or the hyperfine interaction and a portion of the initial $S$ state is transformed in $T_+$ \cite{pettaCoherentManipulationCoupled2005a,bertrandQuantumManipulationTwoElectron2015a,fogartyIntegratedSiliconQubit2018}.
In order to transfer a higher portion of the initial singlet population to $T_+$, we perform an adiabatic transformation of the spin state.
The state transfer is characterized by performing a $V_{\textrm{B2}}$ voltage ramp across the previously localized $S$-$T_+$ avoided crossing at a variable speed $v_{\textrm{ramp}}$.
In Fig.~\ref{fig:tools}(d), we plot the singlet probability as a function of $1/v_{\textrm{ramp}}$, as expected by the Landau-Zener formula the transformation rate follows an exponential decay and yields a state coupling close to \SI{300}{\nano\electronvolt} \cite{zenerNonadiabaticCrossingEnergy1932}.
In the end, the transformation rate fidelity in the operating regime is close to \SI{80}{\percent}.
The transformation rate could be limited by different factors such as charge and magnetic noise or the digitalization of the AWG output which all reduce the adiabaticity of the pulse \cite{qiEffectsChargeNoise2017a,nicholQuenchingDynamicNuclear2015}.
This operation being reversible, a $T_+$ state can be transformed selectively into $S$ with the same transformation rate.

The results presented in the two last sections demonstrates our capability to perform on-demand selective transformation of a $T_0$, $T_+$ state into $S$ via respectively, a relaxation hotspot and adiabatic passage through the avoided crossing.
These two key features, associated with the FPSB procedure are at the core of the complete readout procedure that we will develop in the following section.

%< Complete readout
\section{Complete two-electron spin state readout}

In addition to a higher readout fidelity, the capability to conserve the initial spin state at the measurement position allowed us to engineer a readout protocol maximizing the quantity of information extracted from the system.
Indeed, a parity or PSB measurement yields only one bit of information despite having a system containing two qubits.
In this section we develop a protocol entitled complete readout which allows us to discriminate between the four spin states $S$, $T_0$, $T_+$ and $T_-$.
The measurement is based on the repeatable capability of the FPSB allowing us to perform a sequence of three distinct manipulations and measurements to determine the two-electron spin state.

%>> Figures/full readout
\begin{figure*}[h!]
	\centering
	\includegraphics[width=\doublecolumn]{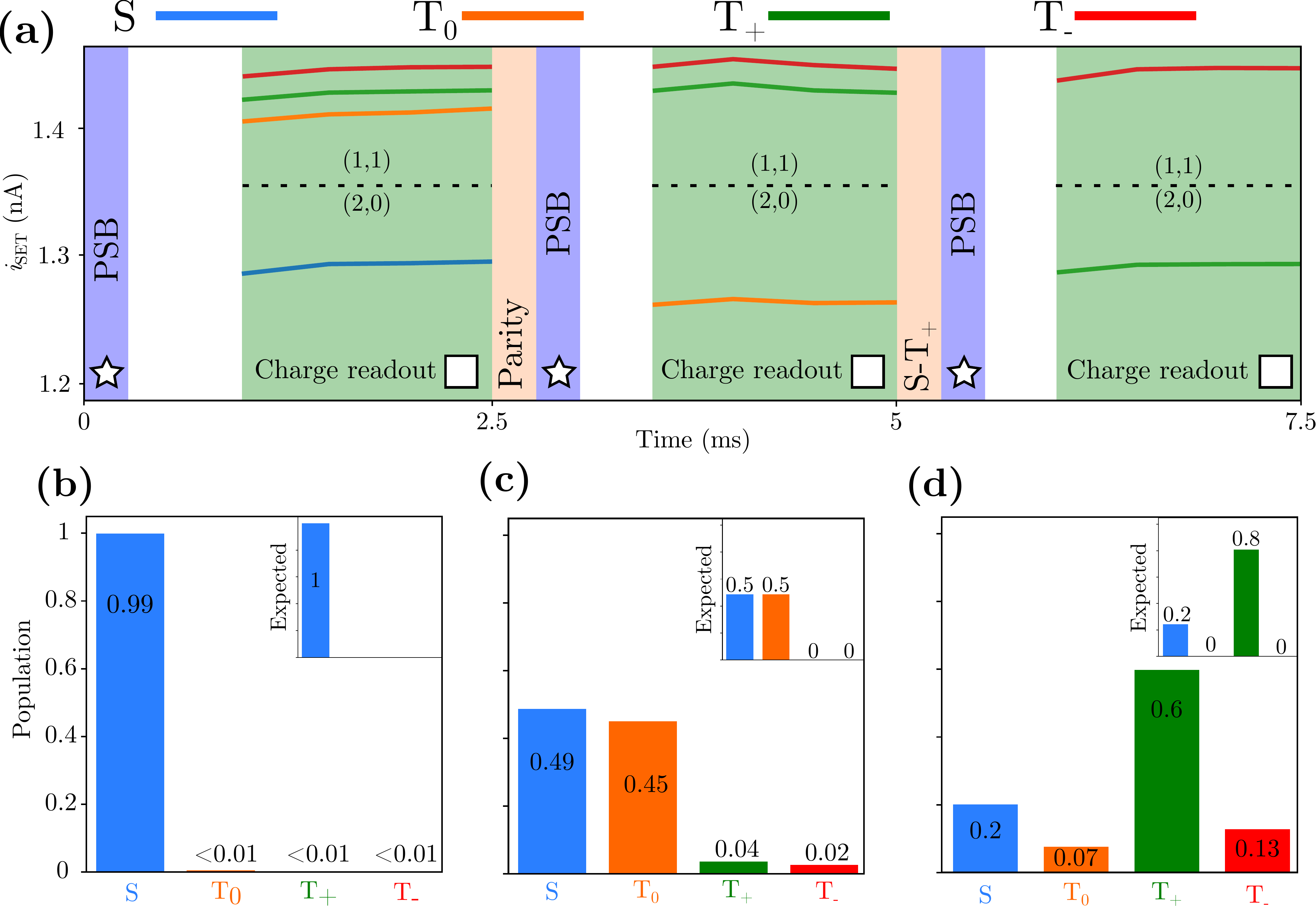}
	\caption{
		\textbf{Complete readout procedure and description of the three-step measurement.}
		(a)~Typical measured current across the SET during the three different steps of the readout for the four possible spin states.
		Blue shaded areas corresponds to a pulse to the PSB position, the green shaded one is when the system is at charge readout position.
		The PSB, parity and $S$-$T_+$ shaded area are artificially extended in the chronogram and the SET traces are slightly offset for clarity.
		The white area corresponds to the time needed for the system to stabilize for readout after a manipulation.
		(b),(c),(d)~Output of the complete readout procedure for singlet state, $S$-$T_0$ mixed state and $T_+$ initialization.
		The inset corresponds to the expected result for an errorless complete readout procedure.
	}
	\label{fig:full-readout}
\end{figure*}
%<< Figures/full readout

In Fig.~\ref{fig:full-readout}(a) we detail the complete readout procedure by plotting the typical signal of the local electrometer during the readout for the four possible spin states.
The first step of the complete readout consists in a FPSB measurement performed as described in the first section.
If the output is (0,2) the shot is attributed to a $S$ no matter what the output of the following measurement is.
If the output is (1,1) the state is one of the three triplet states and we have to consider the two following measurements.
During the second step, we perform a parity readout to discriminate $T_0$  from $T_+$ and $T_-$.
Again the charge configuration is read and a (0,2) is associated to $T_0$ while (1,1) is either a $T_+$ or a $T_-$ state and the last measurement needs to be taken into account.
To discriminate the last two spin states an adiabatic passage is performed through the $S$-$T_+$ avoided crossing.
After this passage the spin state is read using a frozen PSB measurement and a (0,2) output is attributed to $T_+$ since the transformation is selective.
Finally, the (1,1) output is attributed to $T_-$.
The attribution of the \num{8} possible outcomes of the three-step measurement is presented in Table~\ref{table-outcome}.

\begin{table}
	\begin{ruledtabular}
		\begin{tabular}{cccccccc}
			                          &                   &               & \textbf{Initialization} &                \\
			                          & Readout output    & S             & Mixed $S$-$T_0$         & Ramp $S$-$T_+$ \\
			\hline
			\multirow{4}{*}{\singlet} & (2,0)/(2,0)/(2,0) & 0.24          & 0.09                    & 0.03           \\
			                          & (2,0)/(2,0)/(1,1) & \textbf{0.76} & \textbf{0.4}            & 0.14           \\
			                          & (2,0)/(1,1)/(2,0) & $<0.01$       & $<0.01$                 & 0.02           \\
			                          & (2,0)/(1,1)/(1,1) & $<0.01$       & $<0.01$                 & $<0.01$        \\
			\hline
			\multirow{2}{*}{\tzero}   & (1,1)/(2,0)/(2,0) & $<0.01$       & 0.08                    & 0.02           \\
			                          & (1,1)/(2,0)/(1,1) & $<0.01$       & \textbf{0.37}           & 0.05           \\
			\hline
			\multirow{1}{*}{\tplus}   & (1,1)/(1,1)/(2,0) & $<0.01$       & 0.04                    & \textbf{0.6}   \\
			\hline
			\multirow{1}{*}{\tminus}  & (1,1)/(1,1)/(1,1) & $<0.01$       & 0.027                   & 0.13           \\
		\end{tabular}
	\end{ruledtabular}
	\caption{
		\textbf{Outcome probability of all the measurements possible for the three initialization tested.}
		For each initialization the measurement procedure is repeated \num{10000} times.
		All the charge readout output are associated to one of the four spin state.
		The highlighted outputs are the one to be maximized if the complete readout procedure was errorless.
	}
	\label{table-outcome}
\end{table}

In Fig.~\ref{fig:full-readout}(b),(c),(d), the complete readout procedure is implemented for three different initializations to test its validity.
The first one consists in waiting for the relaxation to the $S$ ground state during \SI{5}{\ms} in (0,2) charge region.
We obtain \SI{99.7(1)}{\percent} $S$ population in agreement with the PSB readout fidelity obtained previously.
To test the ability of the complete readout to discriminate $T_0$ state efficiently, we initialize a $S$-$T_0$ mixed state and again apply the readout procedure to it.
The result is shown in Fig.~\ref{fig:full-readout}(b), we observed an excess of $S$ and $T_0$ close to the expected values.
For $T_0$, we obtained \SI{45}{\percent} probability instead of \SI{50}{\percent} this discrepancy is attributed to the limited fidelity of the parity readout (see Appendix \ref{parity-readout-fidelity}).
Finally, we initialized a \SI{79.9(4)}{\percent} population of $T_+$ via an adiabatic ramp through the anticrossing as presented in the previous section.
The result of the readout procedure is plotted in Fig.~\ref{fig:full-readout}(c) and we indeed observed an excess of $T_+$ state indicating again the validity of the readout for this state.
In this situation, the readout is mainly limited by the transformation rate of a $S$ into a $T_+$ and vice versa.
Due to this limited transformation rate \SI{19}{\percent} of the $T_+$ population is mislabelled as $T_-$ resulting in \num{0.13} probability to observe this state.

The repetitive measurements performed on the three different initialized populations allows us to evaluate the fidelity of each stage of the complete readout procedure.
Indeed, by taking into account only the first charge measurement in Table~\ref{table-outcome} for the $S$ and mixed $S$-$T_0$ initializations, we can conduct the same calculations to obtain the $S$-$T$ discrimination fidelity in the complete readout procedure $F_{\textrm{PSB}}^{\textrm{comp.}} = \SI{98.7 \pm 1.1}{\percent}$.
Now looking at the second measurement, corresponding to the parity readout we have $F_{\textrm{parity}}^{\textrm{comp.}} = \SI{93.3 \pm 1.1}{\percent}$ (calculations are detailed in Appendix \ref{parity-readout-fidelity}).
The last measurement and precisely the proportion of (2,0) charge states transformed into (1,1) during the last stage of the complete readout procedure allows us to determine the $T_+$ readout fidelity $F_{\textrm{T}}^{\textrm{comp}} = \SI{87(1)}{\percent}$ limited by the $S$-$T_+$ transformation rate $r = \SI{79.9(4)}{\percent}$.

%>> Figures/full readout
\begin{figure}[h!]
	\centering
	\includegraphics[width=\singlecolumn]{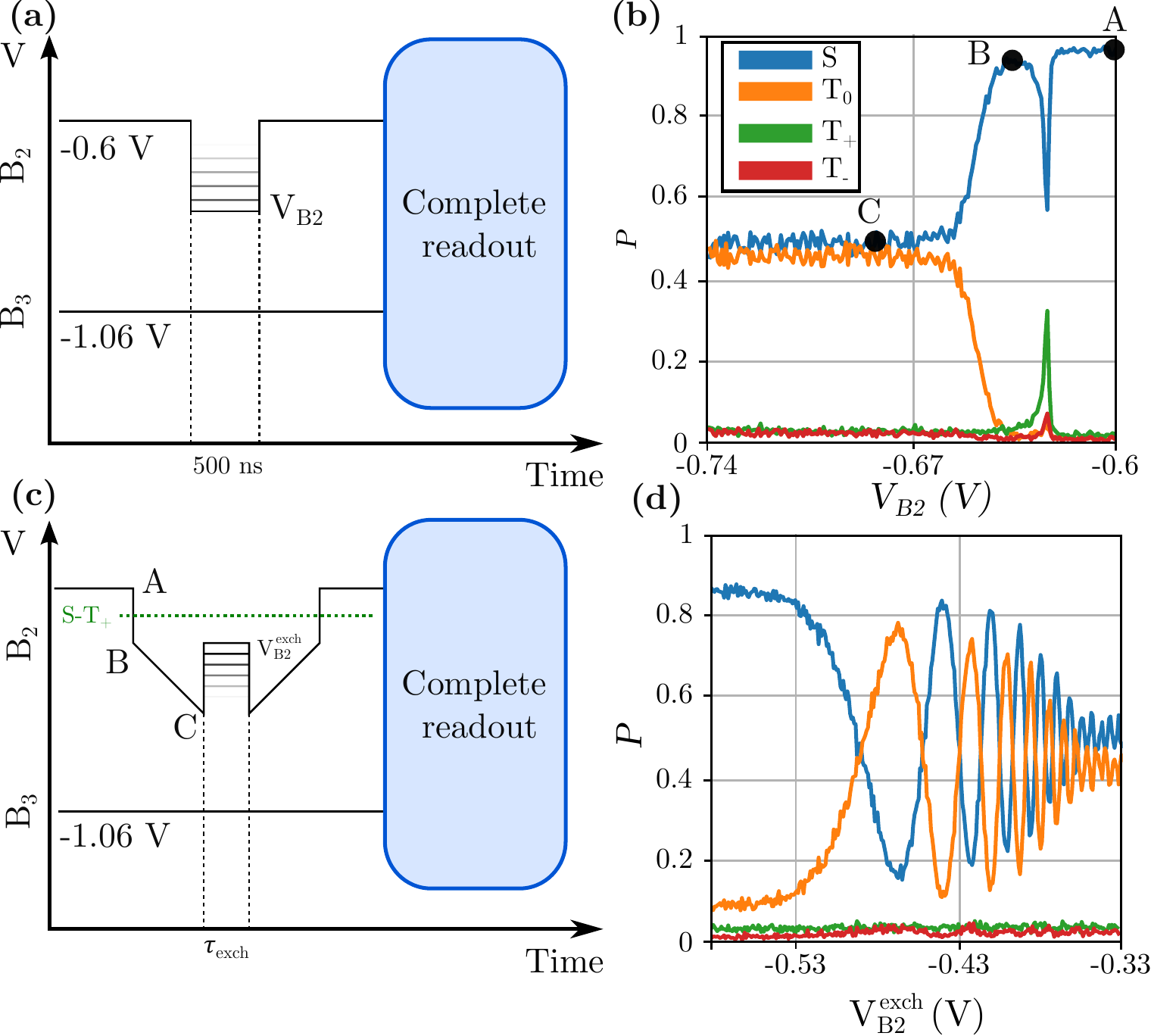}
	\caption{
		\textbf{Spin state operations observed via the complete readout procedure.}
		\textbf(a)~Pulse sequence performed to obtain (b).
		The system is initialized in a singlet state at a large value of $t_c$ ($V_{\textrm{B2}}=\SI[]{-0.6}{\volt}$) in the (1,1) charge state region ($V_{\textrm{B3}}=\SI[]{-1.06}{\volt}$), the inter-dot gate \Btwo is pulsed during \SI{500}{\ns} at a variable amplitude $V_{\textrm{B2}}$.
		A complete readout procedure is then performed, the output is classified using the methodology described in Fig.~\ref{fig:full-readout}.
		(b)~Singlet transformation with respect to the inter-dot tunnel coupling.
		(c)~Pulse sequence used to perform $S$-$T_0$ exchange oscillations.
		The initial \singlet state is conserved from point \textbf{A} to \textbf{B} via a non-adiabatic pulse across the $S$-$T_+$ avoided crossing.
		It is then transformed adiabatically to \updown or \downup (depending on the sign of $\Delta B_z$) via a ramp from \textbf{B} to \textbf{C}.
		From there, the exchange coupling $J$ is increased via an inter-dot tunnel coupling pulse $V_{\textrm{B2}}^{\textrm{exch}}$ during a fixed time $\tau_{\textrm{exch}} = \SI{1.6}{\ns}$.
		During this time the system evolves at a frequency $J$ between \updown and \downup, and finally the mirrored sequence is applied to stop the oscillations and map the two states to $S$ and $T_0$.
		(d)~Spin state probability at the end of the exchange pulse sequence.
		As expected the $S$ and $T_0$ probability are oscillating out of phase.
		We observe a negligible leakage to the $T_+$ and $T_-$ state proving the efficiency of the procedure to perform a SWAP operation.
	}
	\label{fig:operations}
\end{figure}
%<< Figures/full readout

We finally performed classic spin manipulations to ensure the possibility of the complete readout procedure to be inserted in more complex pulse sequence.
In Fig.~\ref{fig:operations}(a), we performed a simple pulse on the inter-dot tunnel coupling and observed through the scope of the complete readout procedure the evolution of a $S$ state.
For $V_{\textrm{B2}}>\SI{-0.61}{\volt}$, $t_c$ is large enough for the $S$ to remain the ground state, it is therefore conserved during the pulse duration and we observe a high probability of $S$.
Lowering slightly more the tunnel coupling we observe a sharp drop in the $S$ probability associated to an increase in the $T_+$ one, as explained previously at this position the two spin states are degenerated and mixing occurs.
The complete readout procedure confirms the observation made in Fig.~\ref{fig:tools}(c).
For even lower values of $t_c$, the $S$ and $T_0$ probabilities both reach \SI{50}{\percent}, this ensures the observation made at the beginning of the article about the $S$-$T_0$ mixing region.
In this regime we observe an excess of the $T_+$ population (\SI{2.8(6)}{\percent}) associated to the passage through the $S$-$T_+$ avoided crossing being slightly adiabatic.

In Fig.~\ref{fig:operations}(d), we performed exchange controlled SWAP gate at the symmetric operation point to have a strong enhancement of the coherence time \cite{bertrandQuantumManipulationTwoElectron2015a,martinsNoiseSuppressionUsing2016,reedReducedSensitivityCharge2016}.
Observing such basic operation through the scope of the complete readout procedure is of major interest to study state leakage.
Indeed, this whole procedure is error-prone due to the high level of control needed over the shape of the pulses.
As we can see in Fig.~\ref{fig:operations}(c), the $S$-$T_+$ crossing and the exchange pulse needs to be performed non-adiabatically while the basis transformation from $S$-$T_0$ to \updown and \downup must be adiabatic which is experimentally challenging.
As expected the $S$ and $T_0$ probabilities are oscillating out of phase while the $T_+$ and $T_-$ ones are remaining low thanks to an efficient non-adiabatic pulse.
However, the oscillations amplitude, even for low exchange coupling, does not exceed \SI{60}{\percent}, the complete readout procedure indicates us that the main limitation comes from the adiabatic transformation performed from the $S$/$T_0$ basis to \updown/\downup and vice versa.
Indeed, a non-adiabatic pulse of the initial $S$ state in the low tunnel coupling regime would result in state mixing process as described in the beginning of this work, which reduces the overall amplitude of the oscillations amplitude.
This level of understanding of the error source is unlocked by the capability to discern the four spin states.
Indeed, in the same situation a simple PSB readout would have been insufficient to determine the cause of the limited oscillations amplitude by yielding the same signal for a $T_+$ and $T_0$ state.

%<< Conclusion
\section{Conclusion}
In this work, we demonstrated a high fidelity spin readout procedure based on PSB by separating in the voltage gate space the spin to charge conversion position from the charge readout.
To do so, we performed the charge conversion via \si{\ns} AWG pulse orders of magnitude faster than the typical relaxation time observed in the GaAs/AlGaAs qubit platform.
The resulting charge configuration was read at fixed and optimized position where the tunnel coupling was sub-\si{\hertz} to prevent any charge variation of the DQD during the electrometer signal acquisition and therefore errors during the readout due to triplet state relaxation.
This novel readout protocol allowed a clear improvement of the readout fidelity and opened the door to more complex readout procedures thanks to its repeatability.
Indeed, we were able to interleave a parity readout based on a selective relaxation hotspot of $T_0$ to $S$ and an adiabatic transformation of $T_+$ to $S$ in between three sequential FPSB readout to perform a complete state discrimination of the two-electron spin states ($S$, $T_0$, $T_+$ and $T_-$).
We confronted the complete readout procedure to various spin initializations to assess its validity and finally implemented it after inducing spin dynamics thanks to the exchange energy control via the interdot tunnel coupling.

In this version of the readout protocol the measurement speed is highly limited by the charge readout step performed using DC measurements of the SET current.
Indeed, all the spin manipulation presented including the relaxation process requires at maximum a pulse length of \SI{5}{\micro\second}.
This limitation could be easily lifted by implementing an RF-SET type of readout allowing to discriminate charge states of a DQD in a few hundreds of \si{\ns} \cite{connorsRapidHighFidelitySpinState2020a}.
Moreover, to improve the overall fidelity of the complete readout each individual spin manipulation needs to be carefully calibrated and optimized.
For the parity readout to reach fidelity closer to the one of the PSB, a deeper understanding of the pulse adiabaticity and state relaxation is required but beyond the scope of this work (see Appendix \ref{parity-calibration}).
Following this, the limited $S$-$T_+$ transformation rate is also one of the biggest source of errors in the readout procedure.
However, the engineering of large and controlled spin-orbit interaction via electric or magnetic control has been demonstrated in various spin qubit platforms and could be a solution to increase the transformation rate \cite{nicholQuenchingDynamicNuclear2015}.

%>>

%<< Acknoledgements
\section*{Acknowledgements}
The authors declare no competing interests.

We acknowledge technical support from the Pole groups of the Institut Néel, and in particular the NANOFAB team who helped with the sample realization, as well as T. Crozes, E. Eyraud, D. Lepoittevin, C. Hoarau and C. Guttin.
This work is supported by the ERC QUCUBE.
A.L. and A.D.W. acknowledge gratefully support of DFG-TRR160 and DFH/UFA CDFA-05-06, DFG project 383065199, and BMBF QR.X Project 16KISQ009.
%>>

\appendix

\section{Sample and experimental setup}

\begin{figure}[h]
	\centering
	\includegraphics[width=40mm]{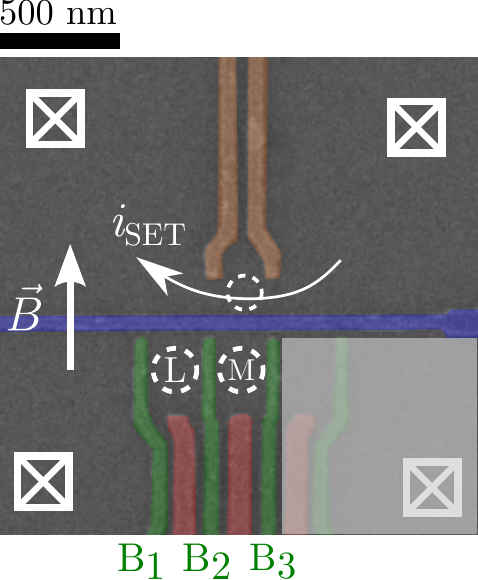}
	\caption{
		Electron micrograph of a sample similar to the one measured.
		The three QDs (lower dashed circles) are probed via a local electrometer (top dashed circle).
		The green gates mainly control the dot-dot and dot-reservoir tunnel rates.
		The red gates are biased with negative voltages to prevent leakage between the gates.
		Four electron reservoirs are located at the white crossed squares and are used to either load electrons in the QD array for the bottom ones or apply a bias through the SET for the top ones.
	}
	\label{fig:SUP-device}
\end{figure}

Our device (presented in Fig.~\ref{fig:SUP-device}) was fabricated using a Si-doped AlGaAs/GaAs heterostructure grown by molecular beam epitaxy, with a two-dimensional electron gas 100~\si{\nano \meter} below the crystal surface which has a carrier mobility of \SI{9e5}{\cm\squared\per\volt\per\second} and an electron density of \SI{2.7e11}{\per\cm\squared}.
Quantum dots are defined and controlled by the application of negative voltages on Ti/Au Schottky gates deposited on top of the heterostructure.

The device is composed of a linear triple QD array, tuned to form a DQD at the positions indicated by the dashed circles L and M.
The grayed part of the array is unused in this article and electron tunneling to this section is prevented by applying negative enough voltages to the gates.
In this configuration $V_{\textrm{B1}}$, $V_{\textrm{B2}}$ and $V_{\textrm{B3}}$ are used to control the reservoir-QD tunnel coupling, the L-M interdot tunnel coupling and the M QD chemical potential.
The charge configuration can be read out by the top quantum dot tuned to be a sensitive local electrometer and biased with 100~\si{\micro\volt}.
The resulting current $\Delta i_{\textrm{SET}}$ is measured using a current-to-voltage converter with a typical bandwidth of 10~\si{\kilo\hertz}.

The gates and ohmic contacts are micro-bonded to a homemade PCB providing DC and RF lines with cut-off frequencies of respectively \SI{100}{\mega\hertz} and \SI{10}{\giga\hertz}.
It is thermally anchored to the mixing chamber of a dry dilution refrigerator with a base temperature of \SI{10}{\milli\kelvin}.
A homemade DAC ensures fast changes of both chemical potentials and tunnel couplings with voltage pulse rise times approaching 100~\si{\nano\second} and refreshed every 6~\si{\micro\second}.
A Tektronix 5014C AWG with a typical channel voltage rise time (\SI{20}{\percent} - \SI{80}{\percent}) of \SI{0.9}{\nano\second} is used to change the $V_\mathrm{B_2}$ and $V_\mathrm{B_3}$ gate voltages at the nanosecond timescale.
These fast gates are connected to \SI{-17}{\decibel} attenuated RF lines and the signal is added to the DAC signal through a homemade bias-tee with a cut-off frequency of \SI{1}{\kilo\hertz}.
The sum of the two signals is computed by assuming that \SI{1}{\volt} applied on the AWG corresponds to \SI{0.1}{\volt} on the gate.

A magnetic field of \SI{500}{\milli\tesla} is applied parallel to the sample to separate in energy the three triplet states $T_0$, $T_+$ and $T_-$.

\section{Pauli spin blockade position calibration}

\begin{figure}[]
	\centering
	\includegraphics[width=\singlecolumn]{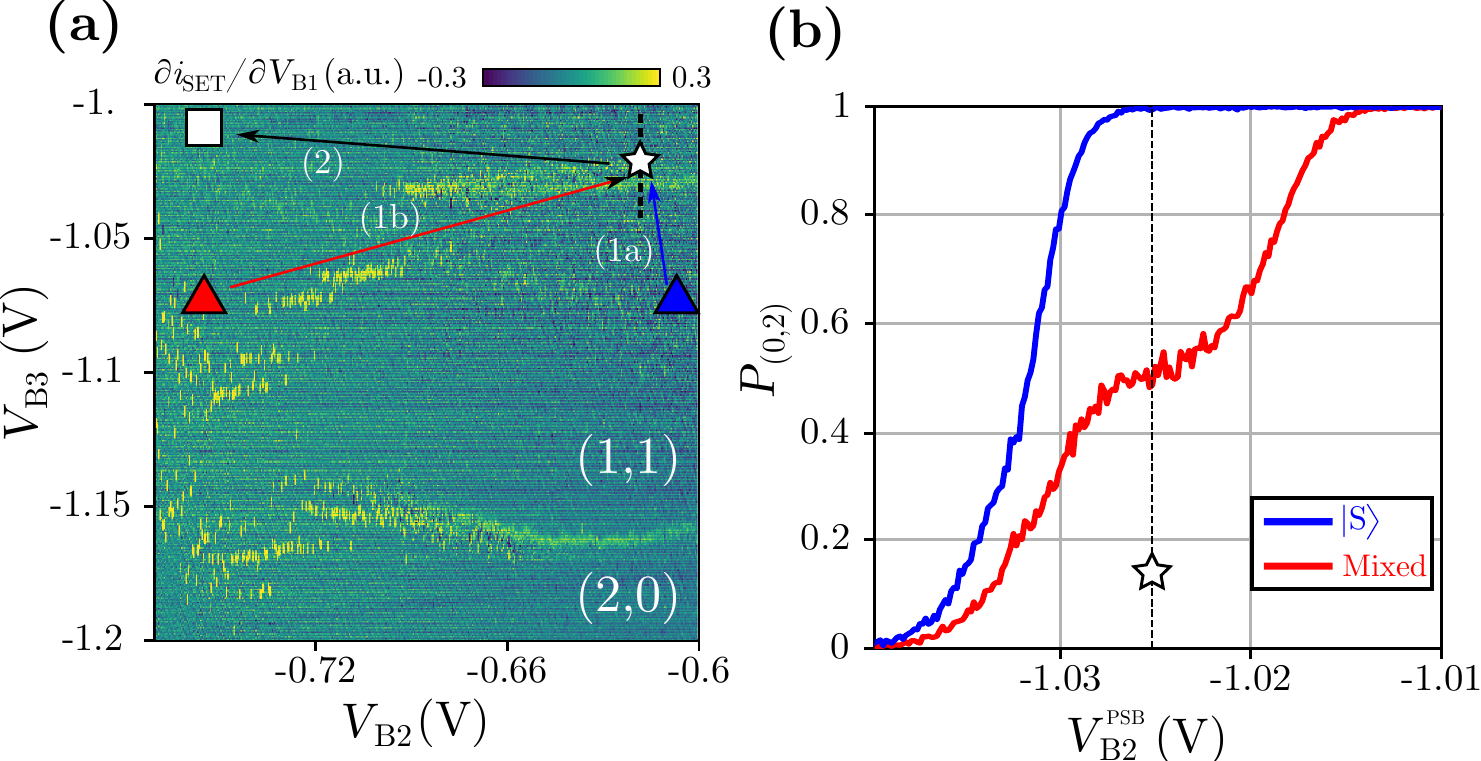}
	\caption{
		\textbf{Calibration of the Pauli spin blockade position in the voltage gate space.}
		(a)~Pulse sequence performed to calibrate the PSB position.
		The system is initialized in a $S$ or mixed $S$-$T_0$ state via a \SI{500}{\ns} voltage pulse at the blue or red triangle respectively.
		It is then pulsed to the PSB position (white star) during \SI[]{20}{\ns} to perform a spin to charge conversion and the configuration of the DQD is read at the white square position during \SI[]{5}{\ms}.
		(b)~PSB calibration experiment.
		The pulse sequence described in (a) is performed for a varied position of the PSB along the detuning axis.
		We plot the (0,2) charge state probability as a function of the PSB position along the dotted line $V_{\textrm{B3}}^{\textrm{PSB}}$ for the two initializations possible.
		The readout position is chosen at the white star position where the mixed state probability reaches \SI[]{50}{\percent}.
	}
	\label{fig:SUP-PSB-calibration}
\end{figure}

To calibrate the PSB position in the voltage gate space, we applied the pulse sequence presented in Fig.~\ref{fig:SUP-PSB-calibration}(a) where we performed a FPSB procedure while varying the PSB position along the black dashed line.
For each position the pulse sequence is repeated \num{10000} times for the two initializations $S$ and $S$-$T_0$ mixed.
We represent in Fig.~\ref{fig:SUP-PSB-calibration}(b), the probability to end up in the (0,2) charge state $P_{(0,2)}$ as a function of the PSB position $V_{\textrm{B3}}^{\textrm{PSB}}$.
For $V_{\textrm{B3}}^{\textrm{PSB}} > \SI{-1.02}{\volt}$, we observe a regime with high probability of (0,2) charge state for both initializations.
Here, the potential detuning overcomes the orbital energy $E_{orb}$ and $S$ and $T$ states have their ground state in the (0,2) charge configuration making it impossible to discriminate.
Similarly, for $V_{\textrm{B3}}^{\textrm{PSB}} < \SI{-1.03}{\volt}$ the two states are in the (1,1) charge configuration yielding low values of $P_{(0,2)}$ for both initializations.
In these two regimes the FPSB readout is not possible, but we observe a region in between where $P_{(0,2)}$ is dependent on the initialization position.
The readout position is selected at the white star, where the $P_{(0,2)} = 0.5$ for the $S$-$T_0$ mixed state initialization which is the theoretically expected result.

\section{Parity readout hotspot localization in the voltage gate space}
\label{parity-calibration}

\begin{figure}[]
	\centering
	\includegraphics[width=\singlecolumn]{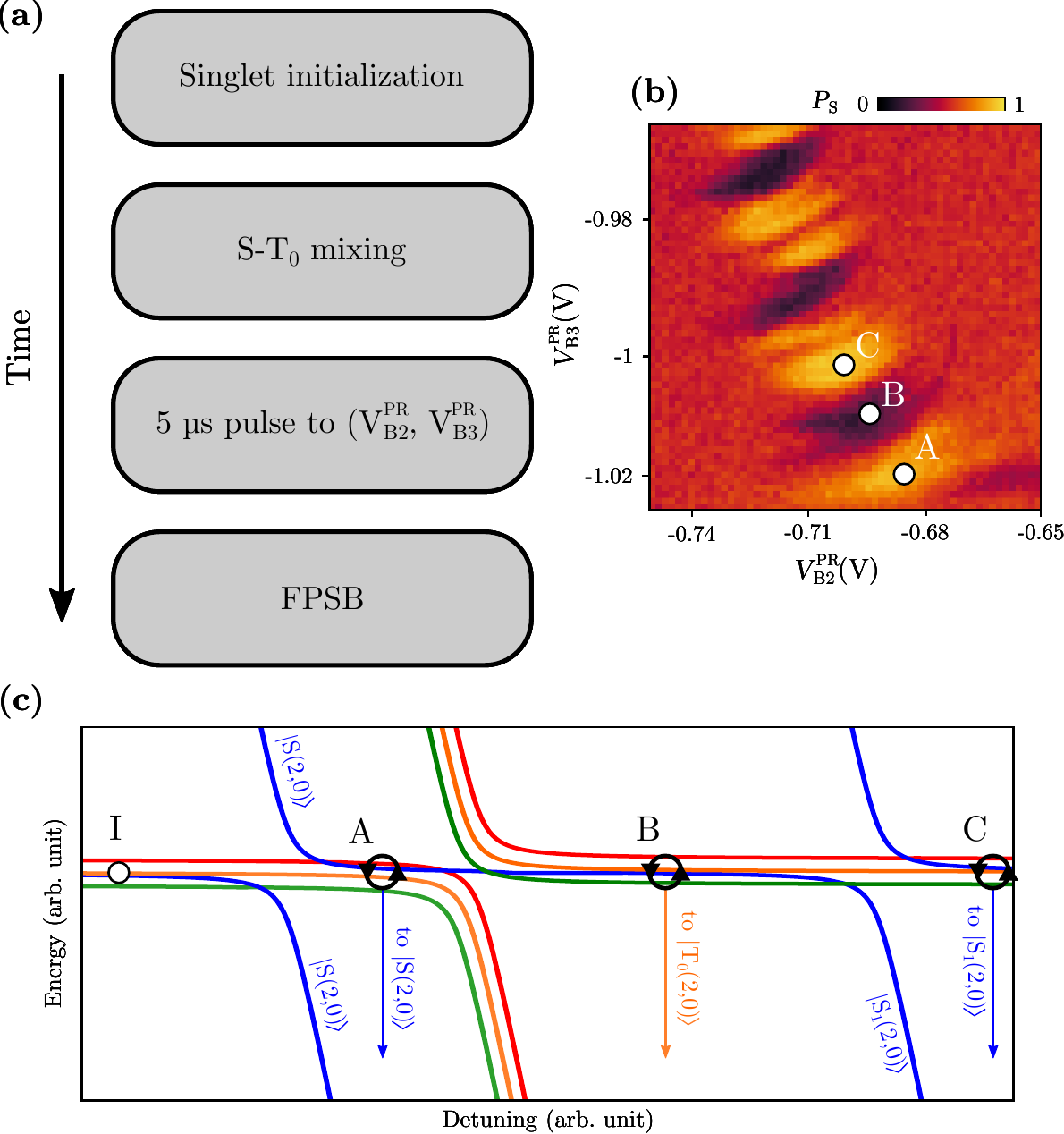}
	\caption{
		\textbf{Calibration of the relaxation hotspot in the voltage gate space.}
		(a)~Experimental workflow used to identify the relaxation hotspot where parity readout is performed.
		The system is initialized in a singlet state and pulsed in the (1,1) charge region at a low value of tunnel coupling in order to prepare an equipartition of $S$ and $T_0$ population.
		The system is then pulsed during \SI{5}{\micro\second} at coordinates ($V_{\textrm{B2}}^{\textrm{PR}}$,$V_{\textrm{B3}}^{\textrm{PR}}$), finally a FPSB procedure is applied to determine the proportion of $S$ at the end of the relaxation pulse.
		(b)~Singlet probability at the end of the pulse sequence presented in \textbf{a} as a function of the pulse coordinates ($V_{\textrm{B2}}^{\textrm{PR}}$,$V_{\textrm{B3}}^{\textrm{PR}}$).
		We observe lobes of high singlet probability indicating that the initialized $T_0$ state has relaxed to the $S$ state during the \SI[]{5}{\micro\second} pulse.
		The position \textbf{C} where the highest singlet probability is recorded, is selected as the parity readout position.
		(c)~Energy diagram of the relevant spin states and identification of the region where \Toneone and \Soneone mixing can occur.
		Once the \Toneone state is transformed it relaxes rapidly through phonon emission to \Sotwoexc.
	}
	\label{fig:SUP-parity-hotspot}
\end{figure}

The parity readout presented in this work relies on a selective relaxation hotspot of $T_0$ to $S$, yielding the same signal for both spin states at the end of the measurement.
The calibration of such hotspot in the voltage gate space is therefore a necessary step to implement the readout.
To do so, we performed the procedure described in Fig.~\ref{fig:SUP-parity-hotspot}(a), where we fully characterized in the $\epsilon$ and $t_c$ parameter space the relaxation of the $S$ and $T_0$ state.

The procedure starts by initializing the spin state in the \Soneone state at a large value of $t_c$, from there it is pulsed during \SI{500}{\ns} at the red triangle position in Fig.~\ref{fig:FigPSB}(a) to initialize a \Soneone-\Toneone mixed state.
The system is then pulsed to the coordinates ($V_{\textrm{B2}}^{\textrm{PR}}$,$V_{\textrm{B3}}^{\textrm{PR}}$) during \SI{5}{\micro\second} to let the spin state evolve.
Finally, a FPSB procedure is performed as described in Sec. \ref{sec:FPSB}, and we plot the $S$ state probability $P_\textrm{S}$ in Fig.~\ref{fig:SUP-parity-hotspot}(b) as a function of the pulse coordinates.
We observe lobes of high and low values of $P_{\textrm{S}}$ alternating as the potential detuning of the DQD increases.

To explain this behavior, we plot in Fig.~\ref{fig:SUP-parity-hotspot}(c) the energy diagram of the relevant states as a function of $\epsilon$.
As mentioned earlier, the system is initialized at position I in the energy diagram ($S$-$T_0$ mixed state), from there it is set at a certain value of detuning at the \si{\ns} timescale.
At this value of $t_c$, we expect the pulse to be non-adiabatic regarding the (1,1)/(0,2) charge transition, therefore the \Soneone and \Toneone states are conserved.
From there, different outcomes are possible depending on the detuning amplitude of the pulse, close to position \textbf{A} the spin states are nearly degenerated and mixing occurs.
As soon as the \Toneone state is transformed into \Soneone it relaxes to the ground state \Sotwo via phonon emission, this behavior corresponds to the bottom lobe of high $P_\textrm{S}$ in Fig.~\ref{fig:SUP-parity-hotspot}(b).
Between position \textbf{A} and \textbf{B}, the \Soneone and \Toneone are split in energy due to the charge state avoided crossing of the $T_0$ state and the initial mixed state is conserved.
When the system is brought close to position \textbf{B} mixing between the state is once again possible, however at this position relaxation is possible to the two state \Sotwo and \Totwo (relaxation to $T_+$ and $T_-$ states is disregarded due to the spin flip process, typically $\simeq \si{\milli\second}$).
Nevertheless, in previous experiments the singlet relaxation rate \Soneone to \Sotwo was found to follow a $\epsilon^{-2}$ dependence \cite{barthelRelaxationReadoutVisibility2012,fujisawaSpontaneousEmissionSpectrum1998} and should therefore be negligible compared to the \Toneone to \Totwo relaxation at this position.
The low $P_\textrm{S}$ lobe around position \textbf{B} arises from the following dynamic: the \Soneone state is transformed into \Toneone via spin mixing which finally rapidly relaxes to \Totwo.
The succession of high and low probability lobes is finally repeated for higher detuning and involves excited charge states of the DQD system.
For instance at position \textbf{C}, mixing between \Soneone and \Toneone is again possible and relaxation to \Sotwoexc state ($S$ state involving an electron on the first excited level of the QD) is the dominant one yielding at the end of the procedure a high $P_{\textrm{S}}$ value.

The position retained at the end of the calibration is the one yielding the highest singlet return probability in a \SI{5}{\micro\second} here point \textbf{C} in Fig.~\ref{fig:SUP-parity-hotspot}(b).

\section{Parity and $T_+$ readout fidelity}
\label{parity-readout-fidelity}

To assess the fidelity of the parity readout, we define the errors $S^{\textrm{p}}_{\textrm{error}}$, $T_{0,\textrm{error}}^{\textrm{p}}$ the probability to measure an even state when the system is initialized in $S$ and $T_0$ respectively.
Similarly, $T_{+,\textrm{error}}^{\textrm{p}}$ and $T_{-,\textrm{error}}^{\textrm{p}}$ are the probabilities to measure an odd state when the system is in the $T_+$ and $T_-$ states.
To compute each contribution in the infidelity of the parity readout we have the ability to initialize a $S$ state, a mixed $S$-$T_0$ state and a statistical mixture composed of $r$ of $T_+$ and $1-r$ of $S$ where $r$ is the transformation rate of the adiabatic ramp presented in the main text.
We finally relate the measurement errors to the outcome through the following equations :

\begin{eqnarray}
	P_{odd}^{\textrm{S}} &=& 1-S^{\textrm{p}}_{\textrm{error}}\;,
	\\
	P_{odd}^{\textrm{mixed}} &=& 1-\frac{S^{\textrm{p}}_{\textrm{error}} + T_{0,\textrm{error}}^{\textrm{p}}}{2} \;,
	\\
	P_{odd}^{\textrm{ramp}} &=& (1-r) \left(1-S^{\textrm{p}}_{\textrm{error}}\right) + r  T_{+,\textrm{error}}^{\textrm{p}}\;.
\end{eqnarray}

Where $P_{odd}^{\textrm{S}}$, $P_{odd}^{\textrm{mixed}}$ and $P_{odd}^{\textrm{ramp}}$ is the odd spin state probability when the system is initialized in respectively $S$, $S$-$T_0$ and $S$-$T_+$.
We define $S^{\textrm{p}}_{\textrm{error}}$, $T_{0,\textrm{error}}^{\textrm{p}}$, $T_{+,\textrm{error}}^{\textrm{p}}$ and $T_{-,\textrm{error}}^{\textrm{p}}$ as the readout infidelity for a $S$, $T_0$, $T_+$ and $T_-$ state.
The three populations of odd state are obtained by summing the probabilities to obtain a (0,2) charge state measurement during the parity stage in the complete readout protocol.
Doing so, allows us to evaluate the fidelity of the parity readout embedded in the complete procedure.
The $S$-$T_+$ transformation rate $r = \SI{79.9(4)}{\percent}$ is obtained by summing the probability to obtain a (1,1) readout at the first measurement in the complete readout procedure when the system is initialized via the adiabatic ramp.
In this case we are assuming that the adiabatic ramp does not create any spin state different from $S$ and $T_+$ and therefore any (1,1) measurement at the $S$-$T$ readout stage is obligatorily a $T_+$.

Solving the system of equations by using the data presented in Table~\ref{table-outcome} we obtain $S^{\textrm{p}}_{\textrm{error}} = \SI{0.4(1)}{\percent}$, $T_{0,\textrm{error}}^{\textrm{p}} = \SI{13.2 \pm 1.9}{\percent}$ and $T_{+,\textrm{error}}^{\textrm{p}}  = \SI{6.5(6)}{\percent}$.
Unfortunately, here we have no way to evaluate $T_{-,\textrm{error}}^{\textrm{p}}$ since we cannot initialize it, but its dynamics should be close to the one of the $T_+$ state and therefore have an equivalent error rate.
We define the overall parity readout fidelity embedded in the complete readout as $F_{\textrm{parity}}^{\textrm{comp}} = 1-(S^{\textrm{p}}_{\textrm{error}}+T_{0,\textrm{error}}^{\textrm{p}}+T_{+,\textrm{error}}^{\textrm{p}}+T_{-,\textrm{error}}^{\textrm{p}})/4 = \SI{93.3 \pm 1.1}{\percent}$.
The infidelity difference between the $S$ and $T_0$ state is attributed to an uncontrolled relaxation of the \Toneone state to \Totwo before the mixing process transforms it to \Soneone at the parity readout position (see Fig.~\ref{fig:SUP-parity-hotspot}).
For $T_{+,\textrm{error}}^{\textrm{p}}$, such a high value could be explained by initialization errors that we cannot distinguish from the measurement ones in our experiment.
Indeed, an undesired creation of $T_0$ state during the initialization of $S$-$T_+$ via the adiabatic ramp would reduce the overall fidelity of the $T_+$ state during the parity readout since it cannot be distinguished from a $T_0$ during the first stage of the complete readout procedure.

Similarly, we compute the $T_+$ readout fidelity using the following equations :

\begin{eqnarray}
	P_{T+}^{\textrm{S}} &=& S^{\textrm{T}}_{\textrm{error}}\;,
	\\
	P_{T+}^{\textrm{ramp}} &=& r(1-T^{\textrm{T}}_{+\textrm{,error}}) + (1-r)S_{\textrm{error}}^T\;.
\end{eqnarray}

Where $P_{T+}^{\textrm{S}}$ and $P_{T+}^{\textrm{ramp}}$ are the probabilities to measure a $T_+$ state in the complete readout procedure when the system is initialized in a $S$ and $S$-$T_+$ respectively.
We define $S^{\textrm{T}}_{\textrm{error}}$, $T^{\textrm{T}}_{+\textrm{,error}}$ the $T_+$ readout infidelity when the system is initialized in a $S$ and $T_+$ state respectively.
Solving the system we obtain $S^{\textrm{T}}_{\textrm{error}} = \SI{1(1)e-3}{\percent}$ and $T^{\textrm{T}}_{+\textrm{,error}} = \SI{25(1)}{\percent}$ mainly limited by the transformation rate of the $S$ tot the $T_+$ state.
In this set of experiment we do not have access to the infidelity when the system is initialized in the $T_0$ and $T_-$ state.
We finally define the fidelity of the $T_+$ measurement as $F_{\textrm{T}}^{\textrm{comp}} = 1-(S^{\textrm{T}}_{\textrm{error}}+T^{\textrm{T}}_{+\textrm{,error}})/2 = \SI{87(1)}{\percent}$.

\bibliography{biblio}

\end{document}